\documentclass[preprint,12pt]{aastex}

\shortauthors{Winn, Holman, \& Roussanova 2006}
\shorttitle{Tres Transits of TrES-1}

\begin{document}

%
\def\ltsima{$\; \buildrel < \over \sim \;$}
\def\lsim{\lower.5ex\hbox{\ltsima}}
\def\gtsima{$\; \buildrel > \over \sim \;$}
\def\gsim{\lower.5ex\hbox{\gtsima}}
                                                                                          
%

\bibliographystyle{apj}

\title{
The Transit Light Curve (TLC) Project.\\
III.~Tres Transits of TrES-1
}

\author{
Joshua N.\ Winn\altaffilmark{1},
Matthew J.\ Holman\altaffilmark{2},
Anna Roussanova\altaffilmark{1}
}

\altaffiltext{1}{Department of Physics, and Kavli Institute for
  Astrophysics and Space Research, Massachusetts Institute of
  Technology, Cambridge, MA 02139}

\altaffiltext{2}{Harvard-Smithsonian Center for Astrophysics, 60
  Garden Street, Cambridge, MA 02138}

\begin{abstract}

  We present $z$ band photometry of three consecutive transits of the
  exoplanet TrES-1, with an accuracy of 0.15\% and a cadence of
  40~seconds. We improve upon estimates of the system parameters,
  finding in particular that the planetary radius is $1.081\pm
  0.029$~$R_{\rm Jup}$ and the stellar radius is $0.811\pm
  0.020$~$R_\odot$. The uncertainties include both the statistical
  error and the systematic error arising from the uncertainty in the
  stellar mass. The transit times are determined to within about 15
  seconds, and allow us to refine the estimate of the mean orbital
  period: $P=3.0300737 \pm 0.0000026$~days. We find no evidence for
  star spots or other irregularities that have been previously
  reported.

\end{abstract}


\section{Introduction}

The passage of a planet in front of its parent star is an occasion for
celebration and for intensive observations. Transits provide a wealth
of information about an exoplanetary system, even when there is little
hope of ever resolving the system with adaptive optics, coronagraphy,
or interferometry (see, e.g., the recent review by Charbonneau et
al.~2006). The dimming events can be recorded photometrically (as
first done by Charbonneau et al.~2000 and Henry et al.~2000) and used
to determine the planetary and stellar radii. Short-term anomalies in
the timing of the transits may betray the presence of moons or other
planets (Holman \& Murray 2005, Agol et al.~2005), and long-term
variations in the light-curve shape should result from orbital
precession (Miralda-Escud\'{e} 2002), although neither effect has yet
been observed. A few stellar absorption lines have been observed to
deepen due to absorption by the planetary atmosphere (Charbonneau et
al.~2002, Vidal-Madjar et al.~2003). There are also anomalous Doppler
shifts due to the Rossiter-McLaughlin effect, which reveal the angle
on the sky between the stellar rotation axis and the orbital axis
(Queloz et al.~2000; Winn et al.~2005,2006; Wolf et
al.~2006). Secondary eclipses---the counterpoints to transits, when
the planet passes behind the star---have not yet been detected at
optical wavelengths, where the signal would reveal the planetary
albedo. However, they have been detected at mid-infrared wavelengths
(Charbonneau et al.~2005; Deming et al.~2005, 2006) and are beginning
to reveal details of exoplanetary thermal emission.

The aim of the Transit Light Curve (TLC) project is to build a library
of high-precision transit photometry. Our main scientific goals are to
support all of the transit investigations mentioned above, by refining
estimates of the basic system parameters; and to search for secular
and short-term variations in transit times and light-curve shapes that
would be produced by as-yet undiscovered planets or moons. We have
previously reported on TLC observations of the transiting exoplanets
XO-1b (Holman et al.~2006) and OGLE-TR-111b (Winn et al.~2006).

In this paper, we present TLC results for TrES-1b, whose discovery by
Alonso et al.~(2004) was notable for being the first success among the
many ground-based, wide-field surveys for transiting planets with
bright parent stars. The TrES-1 parent star is 12th magnitude with
spectral type K0~V. The planet is a ``hot Jupiter'' with a mass and
radius of approximately 0.8~$M_{\rm Jup}$ and 1.0~$R_{\rm Jup}$, and
an orbital period that is almost exactly 3~days.  Calculations by
Laughlin et al.~(2005) have shown that the measured planetary radius
is in accordance with models of irradiated hot Jupiters.  Charbonneau
et al.~(2005) have detected thermal emission from the planet.  Steffen
\& Agol (2005) have searched for transit timing anomalies, with null
results.

This paper is organized as follows. We describe the observations in
the next section, and the photometric procedure in \S~3. In \S~4, we
describe the techniques we used to estimate the physical and orbital
parameters, and in \S~5 we provide the results, along with a closer
look at the characteristics of the photometric noise. A brief summary
is given in \S~6.

\section{Observations}

We observed three transits of TrES-1 (on UT~2006~June~9, 12, and 15),
corresponding to epochs $E=234$, 235, and 236 of the ephemeris given
by Alonso et al.~(2004):
\begin{equation}
T_c(E) = 2,453,186.8060~{\mathrm{[HJD]}} + E\times(3.030065~{\mathrm{days}}).
\end{equation}
We used KeplerCam on the 1.2m (48~in) telescope at the Fred L.\
Whipple Observatory (FLWO) on Mt.\ Hopkins, Arizona.  KeplerCam has
one 4096$^2$ Fairchild 486 back-illuminated CCD, with a $23\farcm1
\times 23\farcm 1$ field of view. For our observations we used
$2\times 2$ binning, which gives a scale of $0\farcs 68$ per binned
pixel, a readout/setup time of 11~s, and a typical readout noise of
7 e$^-$ per binned pixel.  We observed through the SDSS $z$~band
filter in order to minimize the effect of color-dependent atmospheric
extinction on the relative photometry, and to minimize the effect of
limb-darkening on the transit light curve. The effective band pass was
limited at the blue end by the filter (which has a transmission of
25\% at 825~nm, rising to 97\% for wavelengths redward of 900~nm) and
at the red end by the quantum efficiency of the CCD (which drops from
90\% at 800~nm to 25\% at 1000~nm).

On each of the three nights, we observed TrES-1 for approximately 6~hr
bracketing the predicted midpoint of the transit.  We acquired 30
second exposures in good focus.  We also obtained dome flat exposures
and zero-second (bias) exposures at the beginning and the end of each
night.

On the night of UT~June~9, we observed under generally light clouds,
through an airmass ranging from 1.01 to 1.35.  A period of thicker
clouds caused about 45 minutes of data following egress to be
unusable. There were also small temperature fluctuations
($\sim$0.1~deg~min$^{-1}$) that proved to cause systematic errors in
the photometry, as described in \S~3. In addition, the guider was not
functioning as intended, causing the position of TrES-1 on the
detector to drift by about 20~pixels over the course of the
observations. The full-width at half-maximum (FWHM) of stellar images
ranged from 2 to 3 binned pixels (1\farcs3 to 2\farcs0).

The weather was nearly perfect on the nights of UT~June 12th and 15th,
with no visible clouds.  Automatic guiding was functioning, and the
pixel position of TrES-1 varied by no more than 4 pixels over the
course of each night.  On the 12th, the airmass ranged from 1.01 to
1.18, and the FWHM was nearly constant at approximately 2.4~binned
pixels (1\farcs6).  On the 15th, the airmass ranged from 1.01 to 1.20,
and the FWHM ranged from 2.5 to 3.5~binned pixels (1\farcs7 to
2\farcs3).

\section{Data Reduction}

We used standard IRAF\footnote{
  The Image Reduction and Analysis
  Facility (IRAF) is distributed by the National Optical Astronomy
  Observatories, which are operated by the Association of Universities
  for Research in Astronomy, Inc., under cooperative agreement with
  the National Science Foundation.
}
procedures for the overscan correction, trimming, bias subtraction,
and flat-field division. We performed aperture photometry of TrES-1
and 12 nearby stars of comparable brightness.  Many different aperture
sizes were tried in order to find the one that produced the minimum
noise in the out-of-transit data.  For the three nights, the optimum
aperture radii were 7, 8, and 7~pixels ($4\farcs 7$, $5\farcs 4$, and
$4\farcs 7$).  We subtracted the underlying contribution from the sky,
after estimating its brightness within an annulus centered on each
star ranging from 30 to 35 pixels in radius.

For each night's data, the following steps were followed to produce
relative photometry of TrES-1: (1) The light curve of each comparison
star was normalized to have unit median.  (2) A comparison signal was
created by taking the mean of all 12 normalized light curves, after
rejecting obvious outliers in the comparison star light curves.  (3)
The light curve of TrES-1 was divided by the comparison signal.
Although the transit was obvious, there was a small residual time
gradient in the out-of-transit data, a sign of a residual systematic
error. (4) To correct for the time gradient and set the out-of-transit
flux to unity, the light curve was divided by a linear function of
time. This function was determined as part of the model-fitting
procedure and will be described in the next section.

For the data from UT~June~9, an intermediate step between steps (3)
and (4) was necessary. Those data suffered from an additional
systematic error, namely, a few ``bumps'' in the light curve, with
amplitude $\sim$0.2\% and time scale $\sim$15~minutes.  The deviations
between this light curve and the light curves based on data from the
subsequent nights proved to be well correlated with both the ambient
temperature (especially the temperature measured at the mirror cell),
and the measured ellipticity of stellar images. Presumably the
temperature fluctuations were affecting the focus, which in turn
caused differential leakage of light outside the digital apertures. To
correct for this systematic error, we used the data from the other two
nights to construct a light-curve model (see \S~4), and then found the
residuals between the UT~June~9 data and the model. A linear function
provided a good fit to the correlation between the residuals and
ellipticity; we used this relation to correct each data point based on
the measured ellipticity. We verified that the resulting light curve
had residuals that were uncorrelated with temperature, FWHM, and
ellipticity. On the other nights, neither the temperature nor the
ellipticity varied as much. The data did not have any noticeable
correlations with external variables, and no corrections were applied.

To estimate the relative uncertainties in our photometry, we computed
the quadrature sum of the errors due to the Poisson noise of the stars
(both TrES-1 and the comparison stars), the Poisson noise of the sky
background, the readout noise, and the scintillation noise (as
estimated according to the empirical formulas of Young~1967 and
Dravins et al.~1998). The dominant term is the 0.1\% Poisson noise
from TrES-1. For the UT~June~9 data, an additional error term was
included in the sum, equal to one-half the size of the
ellipticity-based correction.

\begin{figure}[p]
\epsscale{0.9}
\plotone{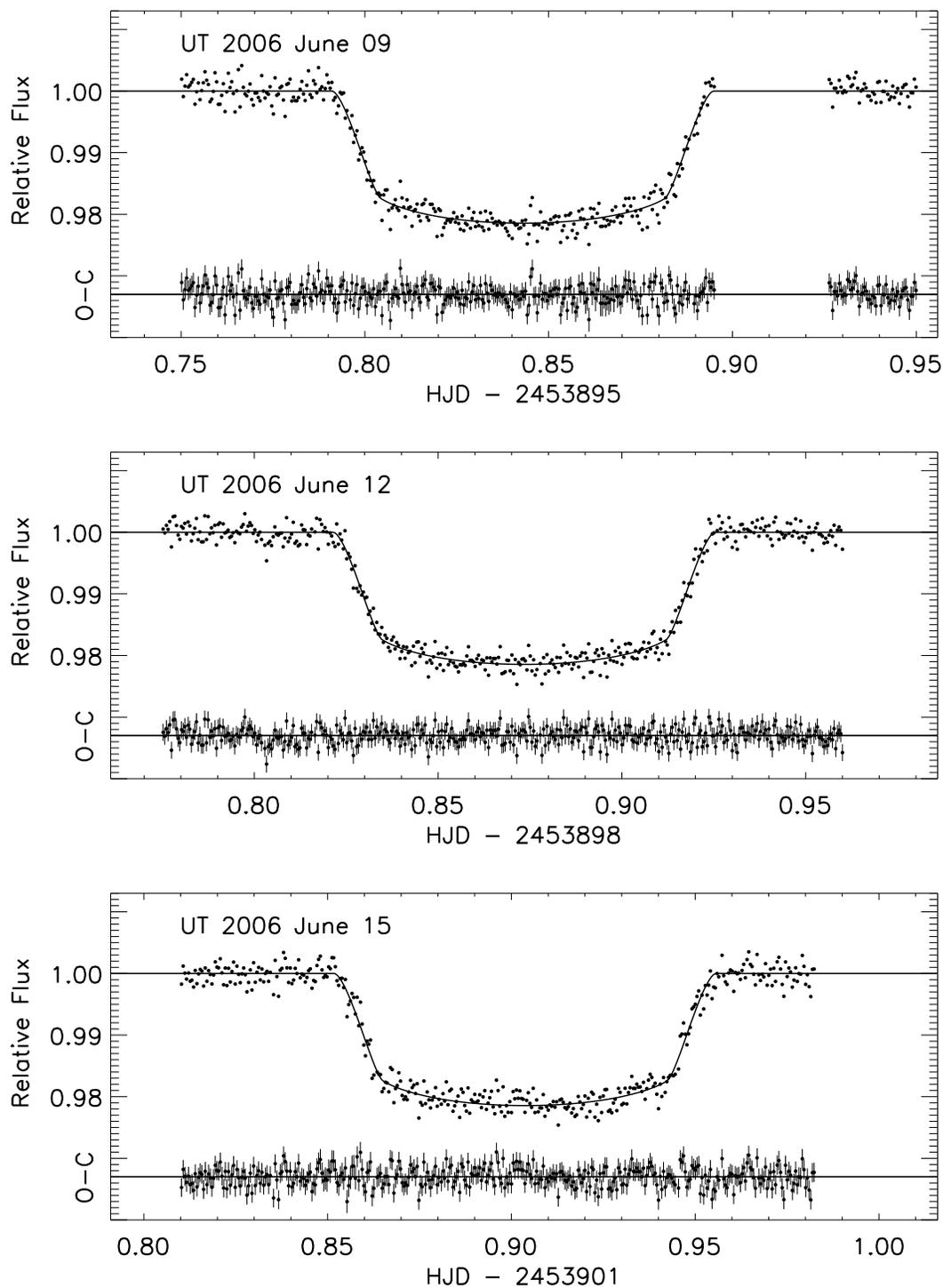}
\caption{Relative $z$ band photometry of TrES-1.  The best-fitting
model is shown as a solid line. The residuals
(observed~$-$~calculated) and the rescaled $1~\sigma$ error bars are also
shown. The residuals have zero mean but are offset
by a constant flux to appear beneath each light curve, for clarity.
From top to bottom, the root-mean-squared (RMS) residuals are 0.16\%, 0.14\%, and 0.15\%.
\label{fig:lc}}
\end{figure}

\begin{figure}[p]
\epsscale{1.0}
\plotone{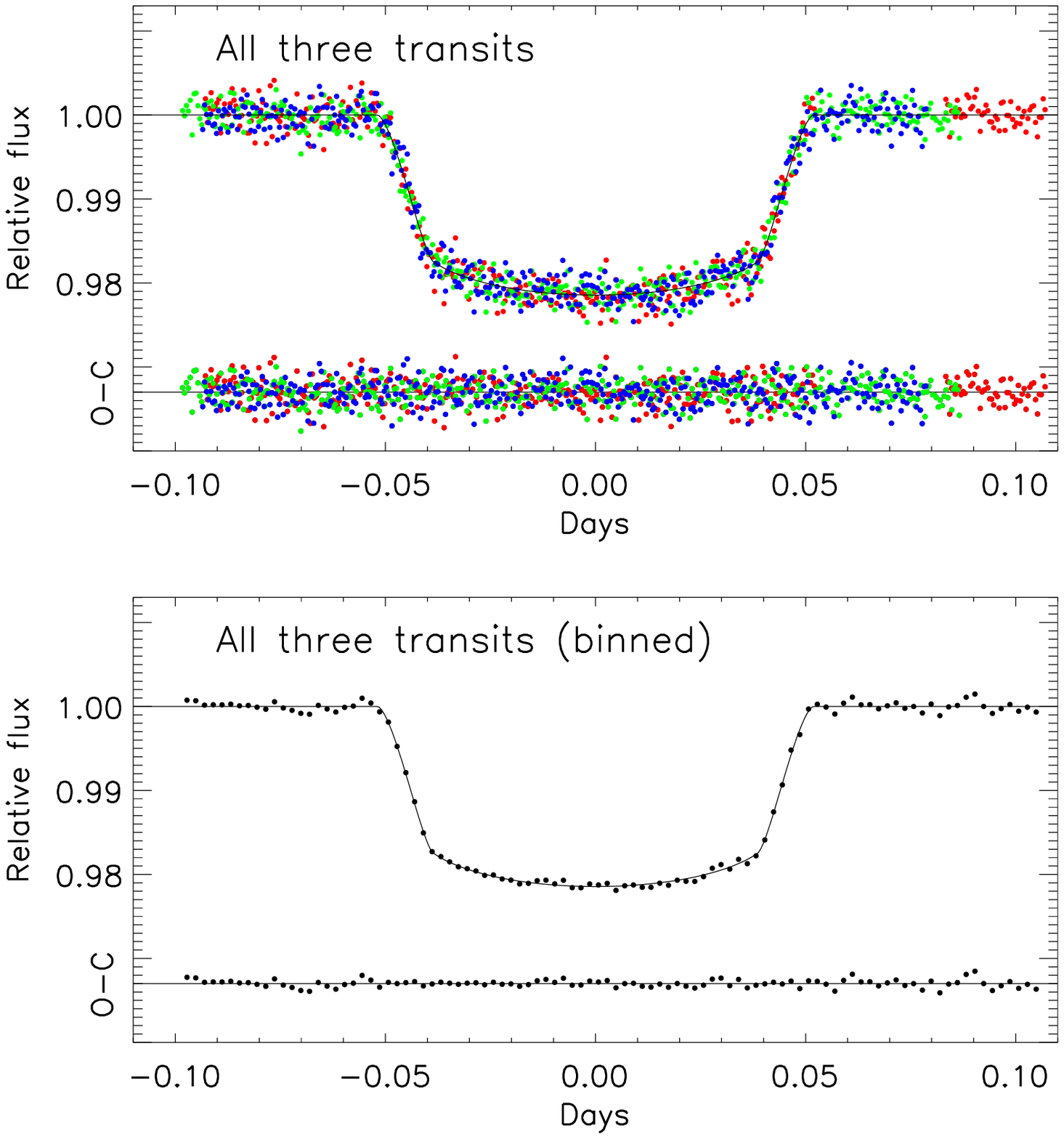}
\caption{Composite transit light curve of TrES-1.  {\bf Top.} Data from
UT~June~9, 12, and 15 are shown with red, green, and blue points,
respectively.  The best-fitting model is shown as a solid line.  The
residuals (observed~$-$~calculated) and the $1~\sigma$ error bars are
shown beneath each light curve. The mean time between points
is 15~seconds, and the RMS residual is 0.15\%. {\bf Bottom.}
Same, after binning in time with 3~minute bins. The RMS residual is
0.046\%.
\label{fig:composite}}
\end{figure}

The final photometry is given in Table~1, and is plotted in
Fig.~\ref{fig:lc}.  The quoted uncertainties have been rescaled by a
factor specific to each night, such that $\chi^2/N_{\rm DOF} = 1$ when
fitting each night's data individually (see \S~4).  The scale factors
by which the calculated errors for each night needed to be increased
were 1.17, 1.13, and 1.26.  A composite light curve, produced by
time-shifting the first and last nights' data appropriately, is shown
in Fig.~\ref{fig:composite}.  The lower panel shows a time-binned
version of the composite light curve, in which the bin size is
3~minutes.

\section{The Model}

To estimate the planetary, stellar, and orbital parameters, we fitted
a parameterized model to all of the photometry simultaneously. The
model is based on a star and a planet on a circular orbit about the
center of mass.\footnote{We assume a circular orbit because, in the
  absence of any evidence for additional bodies in the system, it is
  expected that tidal forces between the star and the planet have had
  sufficient time to circularize the orbit (see, e.g., Rasio et
  al.~1996, Trilling et al.~2000, Dobbs-Dixon et al.~2004).} The star
has a mass $M_\star$ and radius $R_\star$, and the planet has a mass
$M_p$ and radius $R_p$. The orbit has a period $P$ and an inclination
$i$ relative to the sky plane. We define the coordinate system such
that $0\arcdeg \leq i\leq 90\arcdeg$. It is often useful to refer to
the impact parameter $b \equiv a \cos i / R_\star$ (where $a$ is the
semimajor axis) rather than the inclination.

We allowed each transit to have an independent value of $T_c$, the
transit midpoint, rather than forcing them to be separated by exact
multiples of the orbital period.  This is because we sought to measure
or bound any timing anomalies that may indicate the presence of moons
or additional planets in the system. Thus, the period $P$ was relevant
to the model only through the connection between the total mass and
the orbital semimajor axis.  We fixed $P=3.030065$~days, the value
determined by Alonso et al.~(2004).  The uncertainty of
$0.000008$~days was negligible for this purpose (although we were able
to use the resulting values of $T_c$ to refine the period estimate, as
described in \S~5).

The stellar mass cannot be determined from transit photometry alone.
Furthermore, the values of $R_\star$ and $R_p$ that result from
fitting the photometry depend on the choice of stellar mass.\footnote{
  For a fixed $P$, an increase in the stellar mass causes $a$ to
  increase in proportion to $M_\star^{1/3}$. This in turn causes the
  orbital velocity of the planet to be increased and the time scale of
  the transit to be decreased by the same factor. This can be
  compensated for by increasing both $R_\star$ and $R_p$ in proportion
  to $M_\star^{1/3}$, with no other observable effect.} Our approach
was to fix $M_\star$ at a value that has been estimated from an
analysis of the stellar spectrum (its luminosity class, effective
temperature, surface gravity, etc.) and theoretical isochrones, and
then use the scaling relations $R_p \propto M_\star^{1/3}$ and
$R_\star \propto M_\star^{1/3}$ to estimate the systematic error due
to the uncertainty in $M_\star$.  We adopted the value
$M_\star=0.89\pm 0.05$~$M_\odot$ based on the analysis by Sozzetti et
al.~(2004), noting that this value is also in agreement with
independent work by Laughlin et al.~(2005) and Santos et al.~(2006).
The planetary mass $M_p$ is nearly irrelevant to the model (except for
its minuscule effect on the semimajor axis), but for completeness we
use the value $M_p=0.76$~$M_{\rm Jup}$ found by Sozzetti et
al.~(2004).

To calculate the relative flux as a function of the projected
separation of the planet and the star, we assumed the limb darkening
law to be quadratic,
\begin{equation}
\frac{I_\mu}{I_1} = 1 - u_1(1-\mu) - u_2(1-\mu)^2,
\end{equation}
where $I$ is the intensity, and $\mu$ is the cosine of the angle
between the line of sight and the normal to the stellar surface. We
employed the analytic formulas of Mandel \& Agol~(2002) to compute the
integral of the intensity over the portion of the stellar disk hidden
by the planet. There is a quandary over how to treat the limb
darkening parameters when fitting transit photometry. Most previous
investigators in this situation have either fixed the parameters at
the values calculated with stellar atmosphere models (see, e.g., Jha
et al.~2000, Moutou et al.~2004, Laughlin et al.~2005), or allowed
them to vary freely (e.g., Brown et al.~2001, Deeg et al.~2001, Winn
et al.~2005; see also Knutson et al.~2006, who tried both approaches
on the same data). However, the first option is problematic because
the stellar atmosphere models are not perfect, and it may be a mistake
to trust them so completely. On the other hand, the second option
probably allows too much freedom in the limb darkening function, and
hence too much uncertainty in the fitted parameter values.

Our approach was to allow $u_1$ and $u_2$ to be free parameters, but
with a mild {\it a priori} constraint based on the stellar atmosphere
models of Claret~(2004). In particular, we added a penalty term to
$\chi^2$ (see below) to encourage the limb-to-center intensity ratio
$I_0/I_1$ to agree within 20\% of the value calculated by
Claret~(2004) based on the ATLAS quadratic fit ($u_1=0.2851$,
$u_2=0.2849$). We chose the figure of 20\% based on a survey of the
recent literature in which limb darkening calculations are compared to
interferometric and microlensing observations. There seems to be
agreement to within 5-20\% in the limb-to-center intensity ratio for
various stellar types (see, e.g., Abe et al.~2003, Fields et al.~2003,
Ohishi et al.~2004, Aufdenberg et al.~2005, Bigot et al.~2006). We
also investigated the effect of either tightening or dropping this
{\it a priori} constraint, as discussed below.

As noted in the previous section, the light curves exhibited a time
gradient in the out-of-transit data, probably due to differential
extinction between the target star and the comparison stars, or some
other systematic error.  For this reason, each transit was described
with two additional parameters: the out-of-transit flux $f_{\rm oot}$
and a time gradient $\alpha$.

In total, there were 14 adjustable parameters describing 1149
photometric data points. The parameters were $R_\star$, $R_p$, and
$i$; the limb-darkening parameters $u_1$ and $u_2$; and the values of
$T_c$, $f_{\rm oot}$ and $\alpha$ for each of the 3 transits. In
practice, it was preferable to use the parameters $v_1 \equiv 2u_1 +
u_2$ and $v_2 \equiv u_1 - 2u_2$ rather than $u_1$ and $u_2$. This is
because $v_1$ and $v_2$ have nearly uncorrelated uncertainties.

We optimized the parameters by using the AMOEBA algorithm~(Press et
al.~1992, p.\ 408) to minimize the error statistic
\begin{equation}
\chi^2 = \sum_{j=1}^{1149}
\left[
\frac{f_j({\mathrm{obs}}) - f_j({\mathrm{calc}})}{\sigma_j}
\right]^2 + 
\left[
\frac{(I_1/I_0)-0.43}{0.086}
\right]^2
,
\end{equation}
where $f_j$(obs) is the flux observed at time $j$, $\sigma_j$ is the
corresponding uncertainty, and $f_j$(calc) is the calculated value.
The second term is the {\it a priori} constraint on the limb darkening
function. As noted in \S~3, the uncertainties $\sigma_j$ were the
calculated uncertainties (based on Poisson noise, scintillation noise,
etc.), after multiplication by a factor specific to each night, such
that $\chi^2/N_{\rm DOF} = 1$ when each night's data was fitted
individually. (Our intention was not to test the model, but rather to
determine the appropriate weights for the data points.) The optimized
model is plotted as a solid line in Figs.~\ref{fig:lc} and
\ref{fig:composite}. The differences between the observed fluxes and
the calculated fluxes are also shown beneath each light curve.

The statistical uncertainties in the fitted parameters were estimated
using a Markov Chain Monte Carlo (MCMC) simulation (for a brief
introduction, consult appendix A of Tegmark et al.~2004). In this
method, a stochastic process is used to create a sequence of points in
parameter space that approximates the joint probability distribution
of the parameter values, given the data values. The sequence, or
``chain,'' is generated from an initial point by iterating a ``jump
function'' which can take various forms. Our jump function was the
addition of a Gaussian random number to each parameter value. If the
new point has a lower $\chi^2$ than the previous point, the jump is
executed; if not, the jump is only executed with probability
$\exp(-\Delta\chi^2/2)$. We set the perturbation sizes such that
$\sim$25\% of jumps are executed. We created 20 independent chains,
each with 500,000 points, starting from random initial positions.  The
first 100,000 points were not used, to minimize the effect of the
initial condition. The correlation lengths were $\sim$1000 for the
highly covariant parameters $R_p$, $R_\star$, and $b$, and $\sim$200
for the other parameters. The \citet{Gelman.1992} $R$ statistic was
within 0.2\% of unity for each parameter, a sign of good mixing and
convergence.

Table~2 gives the results for each parameter, along with some useful
quantities derived from the parameters. Among the latter are the times
between first and last contact ($t_{\rm IV} - t_{\rm I}$) and between
first and second contact ($t_{\rm II} - t_{\rm I}$), which are useful
for planning future photometric measurements, and the quantity
$(R_p/a)^2$, which controls the amount of starlight that is reflected
from the planet.

\begin{figure}[p]
\epsscale{1.0}
\plotone{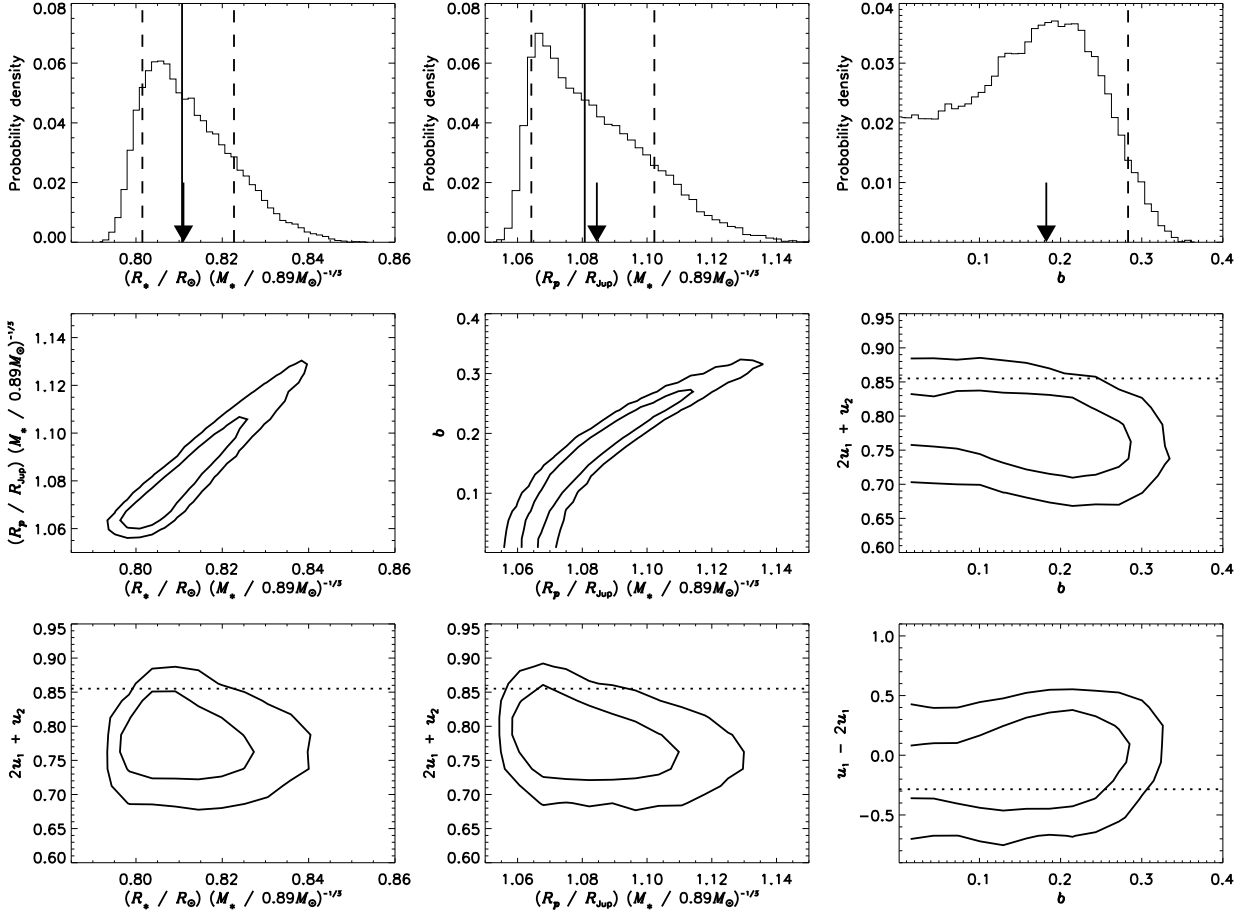}
\caption{
{\bf Top row.} Probability distributions for the stellar radius $R_\star$,
planetary radius $R_p$, and impact parameter $b\equiv a\cos i/R_\star$,
based on the MCMC simulations. The arrows mark the values of the parameters
that minimize $\chi^2$. A solid line marks the median of
each distribution, and the dashed lines mark the 68\% confidence
limits. For $b$, the dashed line marks the 95\% confidence upper limit.
{\bf Middle and bottom rows.}
Joint probability distributions of those parameters with the
strongest correlations.  The contours are isoprobability contours enclosing
68\% and 95\% of the points in the Markov chains.
The dotted lines indicate the values of the limb darkening
parameters calculated by Claret~(2004) using an ATLAS model of a star
with $T_{\rm eff}=5250$~K, $\log g=4.5$, [Fe/H]~=~0.0, and $\xi_t=1.0$~km~s$^{-1}$.
\label{fig:err}}
\end{figure}

Fig.~\ref{fig:err} shows the probability distributions for the
especially interesting parameters $R_\star$, $R_p$ and $b$, along with
some of the 2-d probability distributions for some correlated
parameters. Although the distributions shown in Fig.~\ref{fig:err} are
somewhat asymmetric about the median, Table~2 reports only the median
$p_{\rm med}$ and a single number $\sigma_p$ characterizing the
statistical error. The value of $\sigma_p$ is the average of $|p_{\rm
  med}-p_{\rm hi}|$ and $|p_{\rm med} - p_{\rm lo}|$, where $p_{\rm
  lo}$ and $p_{\rm hi}$ are the lower and upper 68\% confidence
limits.  Two-sided formal uncertainties are of little practical
significance for $R_p$ and $R_\star$ because those parameters are
subject to a systematic error that is comparable to the statistical
error, namely, the covariance with the stellar mass.  The size of this
systematic error is also given in Table~2. The results for $b$ and $i$
are best regarded as one-sided limits, since central transits are
allowed with reasonable probability.

\section{Results}

Our results for the planetary and stellar radii are $R_p = 1.081\pm
0.029$~$R_{\rm Jup}$ and $R_\star = 0.811\pm 0.020$~$R_\odot$. The
quoted error in each parameter is the quadrature sum of the
statistical error (based on the 68\% confidence limits found through
the MCMC simulations) and the systematic error (based on the 1~$\sigma$
uncertainty in the stellar mass quoted by Sozzetti et al.~2004). The
statistical and systematic errors are comparable in size. The ratio of
radii is independent of the stellar mass and is known with greater
precision: $R_p/R_\star = 0.13686 \pm 0.00082$.

Previous investigators have not had photometry of sufficiently high
cadence and signal-to-noise ratio to solve for both $R_\star$ and $R_p$
simultaneously. Instead, they have used estimates for $R_\star$ (as well
as $M_\star$) based on an analysis of the stellar spectrum and theoretical
isochrones. Sozzetti et al.~(2004) found $R_\star = 0.83\pm
0.05$~$R_\odot$, and Laughlin et al.~(2005) found $R_\star = 0.83\pm
0.03$~$R_\odot$.\footnote{
  Laughlin et al.~(2005) also made use of the existing photometry
  showing that the transits are nearly central ($b\approx 0$) to place
  a lower limit of $0.80~R_\odot$ on the stellar radius.
}
Our photometry has allowed us to estimate $R_\star$ independently,
subject only to the fairly weak covariance with $M_\star$. Our result
of $R_\star = 0.811\pm 0.020$~$R_\odot$ confirms the previous
estimates and improves upon their precision.

The planetary radius of $R_p = 1.081 \pm 0.029$~$R_{\rm Jup}$ is also
in agreement with (and more precise than) the previous estimates of
$1.04^{+0.08}_{-0.05}$~$R_{\rm Jup}$ (Sozzetti et al.~2004) and
$1.08\pm 0.05$~$R_{\rm Jup}$ (Laughlin et al.~2005). TrES-1 is an
important case study for theoretical models of the structure of hot
Jupiters, in part because of the comparison to the intensively studied
exoplanet HD~209458. The radius of TrES-1 is smaller by $\sim$20\%,
despite a similar mass and a similar degree of stellar insolation.
Calculations by Laughlin et al.~(2005) and Baraffe et al.~(2005) have
shown that the properties of TrES-1 are generally in line with
theoretical expectations, and that HD~209458 is anomalous. This
provides the important clue that any mechanism to ``inflate''
HD~209458 cannot apply to all hot Jupiters of the same mass. Showman
\& Guillot's~(2002) proposal, that atmospheric circulation patterns
deposit a significant fraction of the stellar insolation into the
planetary interior, provides no apparent reason why TrES-1 and
HD~209458 should be so different, unless perhaps the circulation
patterns depend strongly on surface temperature ($\approx$1060~K for
TrES-1 and 1130~K for HD~209458). Bodenheimer et~al.~(2001) proposed
that HD~209458 is heated internally by eccentricity tides. Winn \&
Holman (2005) proposed that HD~209458 is trapped in a Cassini state
with nonzero obliquity, causing ongoing heating through obliquity
tides. This is naturally a rare occurrence, the implication being that
most hot Jupiters (including TrES-1) have a negligible obliquity.
Adding to the puzzle, two newly discovered exoplanets also appear to
be anomalously large: HAT-P-1 (Bakos et al.~2006) and WASP-1 (Collier
Cameron et al.~2006, Charbonneau et al.~2006).

We also confirm that the transit is fairly central, i.e., the
trajectory of the planet comes close to the center of the stellar
disk. With 95\% confidence, $b < 0.28$ and $i > 88\fdg4$.  All other
things being equal, central transits have the maximum possible
duration and allow for the greatest number of in-transit measurements
(and therefore the greatest signal-to-noise ratio). This makes them
favorable for follow-up photometry. However, they are unfavorable for
certain applications, such as the measurement of orbital precession
through changes in the impact parameter (Miralda-Escud{\'e}~2002), or
of spin-orbit alignment through the Rossiter-McLaughlin effect (Ohta
et al.~2005, Gaudi \& Winn 2006). Specifically, a central transit
provides little leverage on the determination of the angle $\lambda$
between the projected orbital axis and the projected stellar spin
axis, because in that case $\lambda$ is strongly covariant with the
stellar rotation rate.

The fitted values of the limb darkening parameters are $u_1 = 0.28\pm
0.06$ and $u_2 = 0.21\pm 0.12$, with a strong covariance as noted
previously. In comparison, the values calculated by Claret~(2004) are
$u_1 = u_2 = 0.29$. These are in good agreement, which is not
surprising because of the {\it a priori} constraint on the limb
darkening function. If this constraint is dropped, then the optimized
limb darkening parameters are $u_1=0.36$ and $u_2=0.01$ (implying a
limb-to-center intensity ratio of 63\% instead of the theoretical
value of 42\%), and the radius estimates become $R_\star =
0.815$~$R_{\rm Jup}$ and $R_p = 1.096$~$R_{\rm Jup}$. Thus, by
dropping the {\it a priori} constraint, the favored values of the
radii are driven to larger values, but by less than the $1~\sigma$
statistical error determined previously. Conversely, if the limb
darkening parameters are held fixed at the Claret~(2004) values, the
new radius estimates are $R_\star = 0.810$~$R_\odot$ and $R_p =
1.072$~$R_{\rm Jup}$, again within the $1~\sigma$ statistical error of
our quoted results. Limb darkening should therefore be considered as
another source of systematic error, but one that is no larger than
(and is probably smaller than) the systematic error due to the
covariance with stellar mass.

\begin{figure}[t*]
\epsscale{0.8}
\plotone{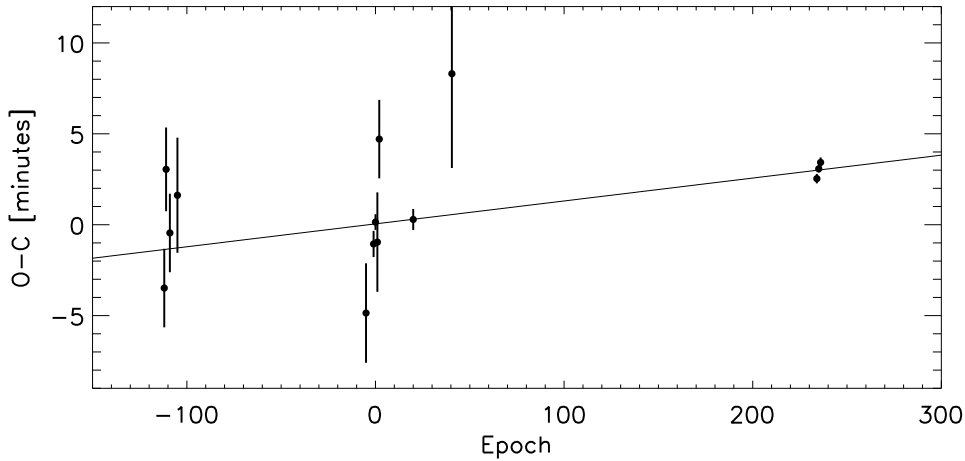}
\caption{ 
Transit and secondary-eclipse timing residuals for TrES-1, from Charbonneau et al.~(2005) and this
work. The calculated times (using the ephemeris of Alonso et
al.~2004) have been subtracted from the observed times.
The best-fitting line is plotted, representing the updated ephemeris
given in Eq.~(\ref{eq:new_eph}) and the text that follows it.
\label{fig:tc}}
\end{figure}

The statistical errors in our transit times range from 12 to
17~seconds. We used the ephemeris of Alonso et al.~(2004) to compute
``observed minus calculated'' (O$-$C) residuals for the newly measured
transit times, along with 14 of the transit times\footnote{We excluded
  the 2 timing measurements that were based on observations of only
  part of the transit, i.e., for which data were missing either prior
  to ingress or after the egress. This is because pre-ingress and
  post-egress data are very important for assessing and correcting
  systematic errors in the photometry.} and the time of secondary
eclipse reported by Charbonneau et al.~(2005). The residuals are shown
in Fig.~\ref{fig:tc}. We fitted a straight line to the transit times
as a function of epoch number $E$ to derive a refined ephemeris
\begin{equation}
T_c(E) = T_c(0) + E\times P,
\label{eq:new_eph}
\end{equation}
finding $T_c(0) =$~2,453,186.80603(28)~[HJD] and $P =
3.0300737(13)$~days, where the numbers in parentheses are the
uncertainties in the last 2 digits. It must be noted that a line is
not a statistically acceptable fit to the timing data; the value of
$\chi^2$ is 23.5 with 12~degrees of freedom ($\chi^2/N_{\rm dof} =
2$).  For this reason, we increased all of the errors by a factor of
$\sqrt{2}$ before determining the uncertainties in $T_c(0)$ and
$P$. The poor fit could be a signal that the period is not constant,
as assumed, or it could be a signal of systematic errors in the
measured times. To be conservative, we recommend basing future
predictions on the period $3.0300737$~days but with an uncertainty of
$0.0000026$~days, i.e., twice as large as the formal
uncertainty. Interestingly, our 3 new measurements are only marginally
consistent with a uniform period at the $1~\sigma$ level. The transits
occur progressively later than expected. Of course, with only 3
measurements this could be a coincidence, but future measurements are
motivated by the possible detection of a satellite or additional
planet through a periodicity in the residuals.

Previous investigators have reported on interesting photometric
anomalies in TrES-1 light curves. The review paper by Charbonneau et
al.~(2006) includes an exquisite light curve obtained with the {\it
  Hubble Space Telescope} in which there is a positive flux excursion
relative to the model of $\approx$0.2\%, lasting for
$\approx$30~minutes (Brown et al., in preparation). The most natural
interpretation is that the planet occulted a star spot or group of
star spots, a phenomenon anticipated by Silva~(2003). We find no
evidence for star spots in our data, and can set bounds on spot
properties under certain assumptions. For a circular spot much smaller
than the planet, the anomaly in the transit light curve caused by the
spot occultation is approximately
\begin{equation}
f(R_{\rm spot}, I_{\rm spot}) = \left( \frac{R_{\rm spot}}{R_\star} \right)^2 \left(1 - \frac{I_{\rm spot}}{I_\star} \right),
\end{equation}
where $R_{\rm spot}$ is the spot radius, $I_{\rm spot}$ is the
intensity of the spot, and $I_\star$ is the intensity of the unspotted
stellar photosphere in the vicinity of the spot. The intensity ratio
can be related to the temperature difference via the blackbody
formula, as in Eq.~(1) of Silva~(2003). The duration of
sunspot-eclipse totality is $\sim$$2R_p/v_{\rm orb}$, or 20~minutes,
if the spot lies exactly on the transit chord. A more likely impact
parameter of 0.5 would lead to a duration of about 15~minutes. By
averaging our data into 15 minute time bins, we find random residuals
with a standard deviation of $\sigma = 2\times 10^{-4}$. Thus we can
set an approximate 2$\sigma$ upper limit of $f < 4\times 10^{-4}$ on
the product of the areal ratio and intensity contrast of small,
circular star spots.  This upper bound is too weak to rule out
Sun-like starspots ($f\lsim 7\times 10^{-6}$),\footnote{ Here, we have
  taken $(R_{\rm spot}/R_\star) \lsim 10^{-5}$, $T_{\rm spot} \approx
  4240$~K and $T_\star \approx 6050$~K based on the typical size and
  temperature contrast of a solar sunspot umbra, from Steinegger et
  al.~(1990). } although the spots on later-type stars like TrES-1
are expected to be more prominent than sunspots (see, e.g.,
Alekseev~2006).  We can rule out flux anomalies as large as the one
seen by Brown et al., but there is no real contradiction because the
observations were taken over a year apart.

In addition, an enterprising group of amateur astronomers created a
composite light curve of TrES-1 transits based on over 10,000
individual brightness measurements (Price et al.~2006).  They reported
statistically significant evidence for a ``brightening episode'' by
0.5\% during egress. We find no evidence for such episodes in any of
our light curves, which have 0.15\% accuracy per 40-second interval.

\begin{figure}[h*]
\epsscale{1.0}
\plotone{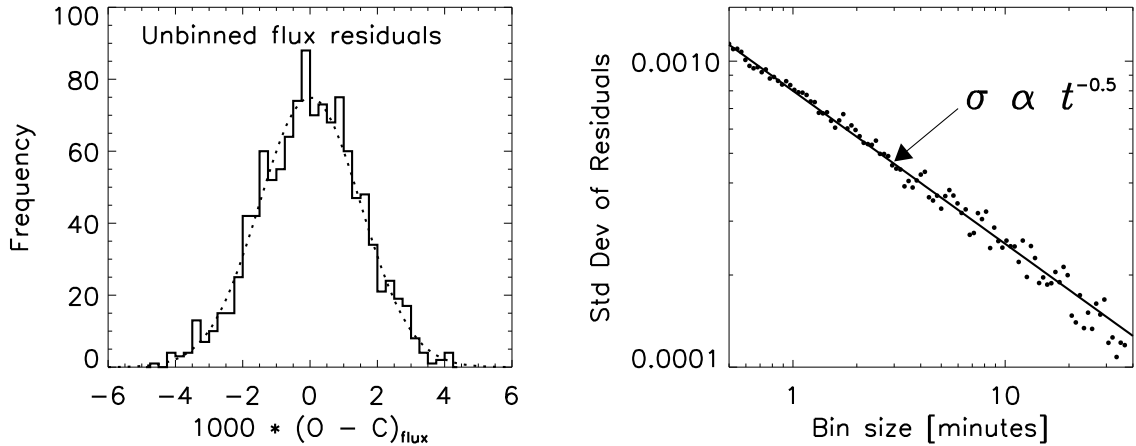}
\caption{
Noise properties of the flux residuals. To compute residuals, the
model-predicted relative flux has been subtracted from each point of
the composite transit light curve shown in Fig.~\ref{fig:composite}. {\bf Left.} The
distribution of unbinned residuals. The dotted line is a Gaussian function with
a standard deviation of 0.15\%. {\bf Right.} The standard deviation of the
residuals, as a function of the size of the time-averaging bin size.
The solid line represents the $1/\sqrt{t}$ dependence that is expected
in the absence of systematic errors.
\label{fig:noise}}
\end{figure}

To characterize our noise statistics more generally, the left panel of
Fig.~\ref{fig:noise} shows a histogram of the flux residuals
(observed~$-$~calculated) from all three nights of data. The
distribution of residuals is approximately Gaussian with a standard
deviation of 0.15\%.  The residuals are not noticeably correlated with
the pixel position of TrES-1, the ambient temperature, or the shape
parameters of stellar images. We also time-averaged the composite
light curve using time bins of various sizes, and calculated the
standard deviation of the residuals as a function of the time bin size
$t$. The results are shown in the right panel of
Fig.~\ref{fig:noise}. The noise is reduced as $1/\sqrt{t}$ over two
orders of magnitude, from the unbinned mean cadence of 15~seconds up
to 30~minutes, at which point we do not have enough data to usefully
average further.

\section{Summary}

Through observations of three consecutive transits, we have
significantly improved upon the estimates of the system parameters for
TrES-1.  Our results are in agreement with previous results, but are
more precise.  We thus confirm the previous conclusions that the
planetary radius is in general agreement with theoretical predictions
for irradiated hot Jupiters. We have also made an independent
determination of $R_\star$ from the photometric signal that confirms
the previous estimates that were based on spectral typing and
theoretical isochrones. We have measured the 3 transit times with an
accuracy of about 15 seconds, leading to a refined estimate of the
mean period, and providing an anchor for continuing searches for
timing anomalies.

The noise in our photometry is nearly Gaussian and averages down with
the expected $1/\sqrt{t}$ dependence all the way to an averaging time
of 30~minutes. We achieved similar results for the exoplanet XO-1
(Holman et al.~2006), although in that case we measured only two
transits with the FLWO 1.2m telescope and Keplercam. We find this
apparent randomness of the residuals to be very encouraging, as it
raises the possibility of detecting signals at the $\sim$10$^{-4}$ or
even smaller levels, such as those produced by smaller transiting
planets, moons, and reflected light. These projects are generally
thought to be the exclusive province of space-based satellite
photometry, but it may be possible to achieve them through extensive
and repeated observations with a relatively small ground-based
telescope.

\acknowledgments We thank Scott Gaudi for very helpful comments about
our analysis and the manuscript. We thank Perry Berlind and Mike
Calkins for help with the observing; Max Tegmark for help with the
MCMC simulations; and Dave Charbonneau, Ben Lane, and Dimitar
Sasselov, for useful discussions. A.R.\ thanks the MIT UROP office for
research funding. Keplercam was developed with partial support from
the Kepler Mission under NASA Cooperative Agreement NCC2-1390
(PI~D.~Latham).

\begin{deluxetable}{lcccc}
\tabletypesize{\normalsize}
\tablecaption{Photometry of TrES-1\label{tbl:photometry}}
\tablewidth{0pt}

\tablehead{
\colhead{HJD} & \colhead{Relative flux} & \colhead{Uncertainty}
}

\startdata
  2453901.96003 &         1.00029 &         0.00154 \\
  2453901.96049 &         0.99700 &         0.00154 \\
  2453901.96094 &         1.00009 &         0.00154 \\
  2453901.96139 &         0.99985 &         0.00154 \\
  2453901.96185 &         0.99998 &         0.00154 \\
  2453901.96230 &         0.99894 &         0.00154 \\
  2453901.96275 &         1.00039 &         0.00154 \\
  2453901.96320 &         1.00139 &         0.00154
\enddata 

\tablecomments{The time stamps represent the Heliocentric Julian Date
  at the time of mid-exposure. The uncertainty estimates are based on
  the procedures described in \S~2. We intend for this Table to appear
  in entirety in the electronic version of the journal. A portion is
  shown here to illustrate its format. The data are also available
  from the authors upon request.}

\end{deluxetable}

\begin{deluxetable}{ccc}
\tabletypesize{\footnotesize}
\tablecaption{System Parameters of TrES-1\label{tbl:params}}
\tablewidth{0pt}

\tablehead{
\colhead{Parameter} & \colhead{Value} & \colhead{Uncertainty}
}

\startdata
                                           $R_\star/R_\odot$& $          0.811$ & $          0.020$\tablenotemark{a} \\
                                           $R_p/R_{\rm Jup}$& $          1.081$ & $          0.029$\tablenotemark{a} \\
             $(R_\star/R_\odot)(M_\star/0.89~M_\odot)^{-1/3}$& $          0.811$ & $          0.012$ \\
             $(R_p/R_{\rm Jup})(M_\star/0.89~M_\odot)^{-1/3}$& $          1.081$ & $          0.021$ \\
                                             $R_p / R_\star$& $        0.13686$ & $        0.00082$ \\
                                                 $R_\star/a$& $         0.0957$ & $         0.0014$ \\
                                                 $(R_p/a)^2$& $       0.000171$ & $       0.000006$ \\
                                                   $i$~[deg]& $         > 88.4$ &    (95\% conf.)\tablenotemark{b} \\  
                                                         $b$& $         < 0.28$ &    (95\% conf.)\tablenotemark{b} \\
                               $t_{\rm IV} - t_{\rm I}$~[hr]& $          2.497$ & $          0.012$ \\
                              $t_{\rm II} - t_{\rm I}$~[min]& $          18.51$ & $           0.63$ \\
                                                       $u_1$& $          0.284$ & $          0.061$ \\
                                                       $u_2$& $           0.21$ & $           0.12$ \\
                                                $2u_1 + u_2$& $          0.777$ & $          0.046$ \\
                                                $u_1 - 2u_2$& $          -0.13$ & $           0.29$ \\
                                           $T_c(234)$~[HJD]& $  2453895.84297$ & $        0.00018$ \\
                                           $T_c(235)$~[HJD]& $  2453898.87341$ & $        0.00014$ \\
                                           $T_c(236)$~[HJD]& $  2453901.90372$ & $        0.00019$
\enddata

\tablecomments{The values in Column 2 are the medians $p_{\rm med}$ of
  the MCMC distributions, which are based on the assumption $M_\star =
  0.89~M_\odot$ and an {\it a priori} constraint on the limb
  darkening function (see the text). Except where noted, the value in
  Column 3 is the average of $|p_{\rm med}-p_{\rm lo}|$ and $|p_{\rm
    med}-p_{\rm hi}|$, where $p_{\rm lo}$ and $p_{\rm hi}$ are the
  lower and upper 68\% confidence limits. (The cumulative probability
  for $p<p_{\rm lo}$ is 16\%, and the cumulative probability for
  $p>p_{\rm hi}$ is 16\%.)}

\tablenotetext{a}{These quantities are affected by a systematic error
  associated with the stellar mass. The quoted uncertainties are the
  quadrature sum of the statistical error and the systematic error
  (estimated under the assumption $M_\star = 0.89\pm 0.05~M_\odot$).}

\tablenotetext{b}{Given the shapes of the probability distributions
  for $b$ and $i$, these results are best regarded as one-sided
  limits.}

\end{deluxetable}

\end{document}